\documentclass[12pt]{article}

\usepackage{amsmath}
\usepackage{amssymb}

\begin{document} 

\begin{titlepage}
\begin{flushright}

\end{flushright}

\begin{center}
{\Large\bf $ $ \\ $ $ \\
A generalization of the Lax pair for the pure spinor superstring 
in $AdS_5\times S^5$
}\\
\bigskip\bigskip\bigskip
{\large Andrei Mikhailov}
\\
\bigskip\bigskip
{\it Instituto de F\'{i}sica Te\'orica, Universidade Estadual Paulista\\
R. Dr. Bento Teobaldo Ferraz 271, 
Bloco II -- Barra Funda\\
CEP:01140-070 -- S\~{a}o Paulo, Brasil\\
}

\vskip 1cm
\end{center}

\begin{abstract}
We show that the Lax pair of the pure spinor superstring in $AdS_5\times S^5$ 
admits a generalization where the generators of the superconformal algebra
are replaced by the generators of some infinite-dimensional Lie superalgebra.
\end{abstract}

\end{titlepage}

\section{Introduction}
\subsection{Lax operator}
The notion of Lax operator is central in the theory of integrable systems. For 
the pure spinor superstring in $AdS_5\times S^5$ the Lax pair was constructed in \cite{Vallilo:2003nx}:
\begin{align}
   L_+  =\;& {\partial\over\partial\tau^+} 
+ \left( J^{[mn]}_{0+} - N^{[mn]}_+ \right)t^0_{[mn]} \;+
\nonumber \\  
&\phantom{\partial\over\partial\tau^+} + {1\over z} J^{\alpha}_{3+}t^3_{\alpha} 
+ {1\over z^2} J^m_{2+}t^2_m + {1\over z^3} J^{\dot{\alpha}}_{1+} t^1_{\dot{\alpha}}
+ {1\over z^4} N^{[mn]}_+t^0_{[mn]}
\label{OrigLPlus}
\\    
L_- = \;& {\partial\over\partial\tau^-} 
+ \left( J^{[mn]}_{0-} - N^{[mn]}_- \right)t^0_{[mn]} \;+
\nonumber \\    
&\phantom{\partial\over\partial\tau^-}
+ zJ^{\dot{\alpha}}_{1-}t^1_{\dot{\alpha}} + z^2 J^m_{2-}t^2_m 
+ z^3J^{\alpha}_{3-}t^3_{\alpha} + z^4N^{[mn]}_- t^0_{[mn]}
\label{OrigLMinus}
\end{align}
where $t^0_{[mn]}, t^1_{\dot{\alpha}}, t^2_m, t^3_{\alpha}$ are generators of ${\bf g} = {\bf psl}(4|4)$: 
\marginpar{${\bf g}$}
\begin{equation}
   {\bf g} = {\bf psl}(4|4)
\end{equation}
satisfying the super-commutation relation\footnote{``Super-commutator'' means commutator
of even with even or even with odd, and anti-commutator of odd with odd 
element; the upper index $\bar{a},\bar{b},\ldots$ of $t^{\bar{a}}_A$ is redundant, it reminds about the ${\bf Z}_4$ grading; the notations we use in this paper
are the same as in \cite{Mikhailov:2007mr} and \cite{Mikhailov:2012uh}} :
\begin{equation}\label{SuperCommutationRelations}
 [t^{\bar{a}}_A,  t^{\bar{b}}_B] = f_{AB}{}^C t^{\bar{a} + \bar{b}\mbox{ \tiny mod } 4}_C
\end{equation}
$J$ are currents: \marginpar{$J$}
\begin{equation}\label{Jvsg}
   J_{\bar{k}} = -(dg g^{-1})_{\bar{k}}
\end{equation}
$z$ is a complex number which is called {\em spectral parameter}.

It follows from the equations of motion that $[L_+,L_-] = 0$. This observation
can be taken as a starting point of the classical integrability theory. 

One can interpret ${1\over z}t_{\alpha}^3$, ${1\over z^2} t^2_m$, ${1\over z^3} t^1_{\dot{\alpha}}$, $\ldots$ as generators of the 
\marginpar{$L{\bf g}$}
twisted loop superalgebra\footnote{The word ``twisted'' means that the power of the spectral
parameter mod 4 should correlate with the ${\bf Z}_4$-grading of the generators} $L{\bf g} = L{\bf psl}(4|4)$. This observation allows us to rewrite (\ref{OrigLPlus}) 
and (\ref{OrigLMinus}) as follows:
\begin{align}
   L_+  =\;& {\partial\over\partial\tau^+} 
+ \left( J^{[mn]}_{0+} - N^{[mn]}_+ \right)T^0_{[mn]} \;+
\nonumber \\  
&\phantom{\partial\over\partial\tau^+} +  J^{\alpha}_{3+}T^{-1}_{\alpha} 
+  J^m_{2+}T^{-2}_m +  J^{\dot{\alpha}}_{1+} T^{-3}_{\dot{\alpha}}
+  N^{[mn]}_+T^{-4}_{[mn]}
\label{LoopLPlus}
\\    
L_- = \;& {\partial\over\partial\tau^-} 
+ \left( J^{[mn]}_{0-} - N^{[mn]}_- \right)T^0_{[mn]} \;+
\nonumber \\    
&\phantom{\partial\over\partial\tau^-}
+ J^{\dot{\alpha}}_{1-}T^1_{\dot{\alpha}} + J^m_{2-}T^2_m 
+ J^{\alpha}_{3-}T^3_{\alpha} + N^{[mn]}_- T^4_{[mn]}
\label{LoopLMinus}
\end{align}
where $T_{\alpha}^{-1}$ replaces ${1\over z}t^3_{\alpha}$ {\it etc.}; operators $T^{n}_A$ are generators of the
twisted loop superalgebra. Withe these new notations, the spectral parameter 
is not present in $L_{\pm}$. Instead of entering explicitly in $L_{\pm}$, it now 
{\em parametrizes a representation} of the generators $T_A^n$.

In this paper we will observe that there is a further generalization. One can
generalize $L{\bf g}$ to the infinite-dimensional superalgebra ${\cal L}_{\rm tot}$ which was
introduced in \cite{Mikhailov:2012uh}.

\subsection{Super-Yang-Mills algebra}
The super-Yang-Mills algebra is the Lie superalgebra formed by the letters $\nabla^L_{\alpha}$
which satisfy:
\begin{equation}\label{IntroPureSpinorConstraint}
   \{\nabla_{\alpha}^L \;,\;\nabla_{\beta}^L\} \Gamma^{\alpha\beta}_{m_1\cdots m_5} = 0
\end{equation}
This is the basic constraint on the covariant derivatives in the $N=4$ $U(N)$
SYM theory:
\begin{equation}\label{IntroCovariantDerivative}
   \nabla^L_{\alpha} = {\partial\over\partial\theta^{\alpha}} + 
   \theta^{\beta}\Gamma^m_{\alpha\beta}{\partial\over\partial x^m} + A_{\alpha}
\end{equation}
where $A_{\alpha} = A_{\alpha}(x,\theta)$ is an $N\times N$ matrix --- the super-vector-potential.
The constraints (\ref{IntroPureSpinorConstraint}) imply all the equations of motion \cite{Witten:1985nt}. 

Let us forget (\ref{IntroCovariantDerivative}) and consider (\ref{IntroPureSpinorConstraint}) as defining an abstract
infinite-dimensional Lie superalgebra. This is called ``SYM superalgebra''. 
This algebra is useful by itself, for example it was conjectured in \cite{Movshev:2009ba} that 
the deformations of this algebra are in one-to-one correspondence with the 
deformations of the maximally supersymmetric Yang-Mills theory. However, in our
opinion, the physical meaning of this algebra remains to be understood. 

Is there a similar algebra for SUGRA? We do not know the answer to this 
question. However, in \cite{Mikhailov:2012uh} we introduced some infinite-dimensional Lie 
superalgebra associated to the Type IIB superstring in $AdS_5\times S^5$. It was 
useful for the study of massless vertex operators. In this paper we will 
explain that this algebra is also naturally related to the structure of the Lax
operator.

\section{Infinite-dimensional Lie superalgebra associated with 
$AdS_5\times S^5$}\label{sec:Lie}
In this section we will discuss the infinite-dimensional Lie superalgebra
${\cal L}_{\rm tot}$ introduced in \cite{Mikhailov:2012uh}. We will show that the definition of the commutator is
self-consistent.

\subsection{SYM superalgebra}
The construction of the SYM superalgebra uses the Koszul duality of quadratic
algebras. Consider the algebra $A = {\bf C}[\lambda_L]$ which is the algebra of polynomial 
functions of the pure spinor $\lambda_L$:
\begin{equation}\label{PureSpinorCone}
   \Gamma^m_{\alpha\beta}\lambda_L^{\alpha}\lambda_L^{\beta}=0
\end{equation}
This is a commutative algebra, therefore the Koszul dual $A^!$ of $A$ is the 
universal enveloping of a Lie superalgebra; this Lie superalgebra is formed 
by the letters $\nabla^L_{\alpha}$ which satisfy:
\begin{equation}\label{PureSpinorConstraint}
   \{\nabla_{\alpha}^L \;,\;\nabla_{\beta}^L\} \Gamma^{\alpha\beta}_{m_1\cdots m_5} = 0
\end{equation}
By definition $A^!$ is the factor-algebra of the tensor algebra, {\it i.e.}
it is formed by linear combinations of expressions of the form:
\begin{equation}
   \nabla_{\alpha_1}\otimes\nabla_{\alpha_2}\otimes\cdots\otimes\nabla_{\alpha_p}
\end{equation}
modulo the subspace formed by the expressions of the form:
\begin{equation}
   \cdots\otimes \Gamma_{m_1\cdots m_5}^{\alpha\beta}\nabla_{\alpha}\otimes\nabla_{\beta}\otimes\cdots
\end{equation}

\subsection{Gluing together two SYM superalgebras}
Consider two copies of the Yang-Mills super algebras, ${\cal L}_L$ generated 
by $\nabla^L_{\alpha}$, and ${\cal L}_R$ generated by $\nabla^R_{\dot{\alpha}}$ (the use of the dotted spinor indices to 
enumerate the generators of ${\cal L}_R$ is traditional). \marginpar{${\cal L}_{L|R}$}
Those are Lie superalgebras, they are generated by all the possible 
super-commutators of $\nabla_{\alpha}^L$ and $\nabla_{\dot{\alpha}}^R$. As in \cite{Mikhailov:2012uh}, we will introduce the Lie 
superalgebra ${\cal L}_{\rm tot}$ which as a linear space is the direct sum: \marginpar{${\cal L}_{\rm tot}$}
\begin{equation}\label{Ltot}
   {\cal L}_{\rm tot} = {\cal L}_L + {\cal L}_R + {\bf g}_{\bar{0}}
\end{equation}
where ${\bf g}_{\bar{0}}$ is a finite-dimensional Lie algebra:\marginpar{${\bf g}_{\bar{0}}$}
\begin{equation}
   {\bf g}_{\bar{0}} = sp(2)\oplus sp(2)
\end{equation}
We will now define the structure of a Lie superalgebra on ${\cal L}_{\rm tot}$ and prove the
consistency of the definition. The basic commutation relations are given by:
\begin{align}
   \{\nabla^L_{\alpha},\nabla^R_{\dot{\beta}}\} =\;& 
   f_{\alpha\dot{\beta}}{}^{[mn]}t^0_{[mn]}
   \label{NablaLNablaR}\\    
[t^0_{[mn]}\;,\;\nabla^L_{\alpha}] = \;&
f_{[mn]\alpha}{}^{\beta}\nabla^L_{\beta}
\label{T0NablaL}\\   
[t^0_{[mn]}\;,\;\nabla^R_{\dot{\alpha}}] = \;&
f_{[mn]\dot{\alpha}}{}^{\dot{\beta}}\nabla^R_{\dot{\beta}}
\label{T0NablaR}\\   
\Gamma_{m_1\ldots m_5}^{\alpha\beta} \{\nabla^L_{\alpha}\;,\;\nabla^L_{\beta}\} = \;& 0
\label{NablaLNablaL}\\   
\Gamma_{m_1\ldots m_5}^{\dot{\alpha}\dot{\beta}} 
\{\nabla^R_{\dot{\alpha}}\;,\;\nabla^R_{\dot{\beta}}\} = \;& 0
\label{NablaRNablaR}
\end{align}
The ${\cal L}_{\rm tot}$ can be considered a universal object in the category of pairs $(\phi,L)$ 
where $L$ is a Lie algebra and $\phi$ a linear map from the linear space formed by 
the letters $\nabla^L,\nabla^R,t^0$ to $L$, consistent with (\ref{NablaLNablaR}) --- (\ref{NablaRNablaR}). In the rest of 
this subsection we give an actual construction of this universal object. As
a linear space, it is (\ref{Ltot}). It remains to define the commutator of two 
arbitrary elements of ${\cal L}_{\rm tot}$ and show that this prescription is compatible with 
the Jacobi identity. 

\subsubsection{Definition of the commutator}
\paragraph     {Subalgebras ${\cal L}_L$ and ${\cal L}_R$}
First of all, let us remember the structure of ${\cal L}_L$ and ${\cal L}_R$. Let $V_L$ denote
the linear space generated by $16$ basis elements $\nabla^L_{\alpha}$, $\alpha\in \{1,\ldots,16\}$. 
Similarly, $V_R$ will stand for the $16$-dimensional linear space generated by
$\nabla^R_{\dot{\alpha}}$. For any linear space $V$, the tensor algebra $T(V)$ is the free 
\marginpar{$T(V)$}
associative algebra generated by the elements of $V$. Its elements are 
denoted, as usual, $v_1\otimes\cdots\otimes v_p$, where $v_i\in V$. Notice that $T(V)$ is a 
graded algebra, the degree being the rank of a tensor. Consider the 
following elements ${\cal R}_{Lm_1\cdots m_5}\in T(V_L)$ and ${\cal R}_{Rm_1\cdots m_5}\in T(V_R)$:
\marginpar{$\langle{\cal R}_{L|R}\rangle$}
\begin{align}
   {\cal R}_{Lm_1\cdots m_5} = &\;      \Gamma_{m_1\cdots m_5}^{\alpha\beta}
\nabla^L_{\alpha}\otimes\nabla^L_{\beta}
\\   
{\cal R}_{Rm_1\cdots m_5} = &\;     \Gamma_{m_1\cdots m_5}^{\dot{\alpha}\dot{\beta}}
\nabla^R_{\dot{\alpha}}\otimes\nabla^R_{\dot{\beta}}
\end{align}
We denote $\langle{\cal R}_L\rangle$ the ideal of $T(V_L)$ generated by ${\cal R}_{Lm_1\cdots m_5}$, and similarly 
$\langle{\cal R}_R\rangle\subset T(V_R)$. Introduce the quadratic algebras: \marginpar{$A_{L|R}$}
\begin{align}
   A_L =\;& T(V_L)/\langle {\cal R}_L \rangle
\\     
A_R =\;& T(V_R)/\langle {\cal R}_R \rangle
\end{align}
It is a general fact about quadratic 
algebras\footnote{Koszul dual to a 
commutative algebra is a universal enveloping of a Lie algebra},  that ${\cal A}_L$ is the universtal 
enveloping of the Yang-Mills algebra ${\cal L}_L$, and ${\cal A}_R$ is the universal enveloping 
of ${\cal L}_R$. The subspace of 
primitive elements of $A_L$ is ${\cal L}^L$, and the subspace of primitive elements
of $A_R$ is ${\cal L}^R$.
Primitive elements are those which can be obtained as nested commutators of the
letters $\nabla^L$ (or $\nabla^R$), for example this one:
\begin{equation}
   \nabla^L_{\alpha}\otimes(\nabla^L_{\beta}\otimes \nabla^L_{\gamma} 
   + \nabla^L_{\gamma}\otimes\nabla^L_{\beta}) -
   (\nabla^L_{\beta}\otimes \nabla^L_{\gamma} 
   + \nabla^L_{\gamma}\otimes\nabla^L_{\beta})\otimes \nabla^L_{\alpha}
\end{equation}
is $[\nabla_{\alpha}^L,\{\nabla_{\beta}^L,\nabla_{\gamma}^L\}]$. We therefore identified the linear space (\ref{Ltot}) as the 
direct sum of three linear spaces:
\begin{itemize}
\item the subspace ${\cal L}_L$ of primitive elements of $A_L = T(V_L)/\langle {\cal R}_L\rangle$
\item the subspace ${\cal L}_R$ of primitive elements of $A_R = T(V_R)/\langle {\cal R}_R\rangle$
\item and ${\bf g}_{\bar{0}}$
\end{itemize}

\paragraph     {Commutators: $[{\bf g}_{\bar{0}}, {\bf g}_{\bar{0}}]$, $[{\bf g}_{\bar{0}}, {\cal L}_L]$ and $[{\bf g}_{\bar{0}}, {\cal L}_R]$}
The definition of the commutator of two elements in ${\bf g}_{\bar{0}}\subset {\cal L}_{\rm tot}$:
\begin{equation}
   [t^0_{[kl]}\;,\;t^0_{[mn]}] = f_{[kl][mn]}{}^{[pq]} t^0_{[pq]}
\end{equation}
where $f_{[kl][mn]}{}^{[pq]}$ are structure constants of ${\bf g}_{\bar{0}}\subset {\bf g}$, as in (\ref{SuperCommutationRelations}). 

The definition of the commutators $[u,x_L]$  and $[u,x_R]$ where $u\in {\bf g}_{\bar{0}}$, $x_L\in {\cal L}_L$
and $x_R\in {\cal L}_R$:
\begin{align}
   \left[
      t^0_{[mn]}\;,\; 
      \nabla_{\alpha_1}^L\otimes \cdots\otimes \nabla_{\alpha_p}^L
   \right] =\;& 
   f_{[mn]\alpha_1}{}^{\alpha'_1}
   \nabla^L_{\alpha'_1}\otimes\cdots\otimes\nabla^L_{\alpha_p}
\;+ 
\nonumber \\    
\;&
+ \ldots + f_{[mn]\alpha_p}{}^{\alpha'_p}   
\nabla^L_{\alpha_1}\otimes\cdots\otimes\nabla^L_{\alpha'_p}
\end{align}
and similar formula for $\left[
      t^0_{[mn]}\;,\; 
      \nabla_{\dot{\alpha}_1}^R\otimes \cdots\otimes \nabla_{\dot{\alpha}_p}^R
   \right]$. Again, $f_{[mn]\alpha}{}^{\beta}$ and $f_{[mn]\dot{\alpha}}{}^{\dot{\beta}}$ 
are the structure constants of ${\bf g}$.

The ${\bf g}_0$-invariance of the subspaces ${\cal R}_L\subset V_L\otimes V_L$ and ${\cal R}_R\subset V_R\otimes V_R$ implies
that $[u, x_L] \in \langle {\cal R}_L \rangle$ when $x_L\in \langle {\cal R}_L\rangle$ ,\hspace{10pt}and that $[u, x_R] \in \langle {\cal R}_R \rangle$ 
when $x_R\in \langle {\cal R}_R\rangle$. This shows that our definition of the commutation of elements
of ${\bf g}_{\bar{0}}$ with elements of ${\cal L}_L$ and ${\cal L}_R$ is correctly defined (respects the 
equivalence relations). 

\paragraph     {Commutator $[{\cal L}_L\;,\;{\cal L}_R]$}  Now we have to define the commutator of 
the type $[x_L,x_R]$ where $x_L\in {\cal L}_L$ and $x_R\in {\cal L}_R$. 

We will use the fact that $x_R$ is a nested commutator of the letters $\nabla_{\dot{\alpha}}^R$, 
therefore it is enough to define the commutator $[x_L, \nabla^R_{\dot{\alpha}}]$ and then define the
commutator $[x_L,x_R]$ by commuting $x_L$ consequtively with the constituents of 
$x_R$, {\it i.e.}:
\begin{equation}
   [x_L, \nabla^R_{\dot{\alpha_1}}\otimes\cdots\otimes\nabla^R_{\dot{\alpha_n}}] =
[\ldots [[x_L,\nabla^R_{\dot{\alpha}_1}],\nabla^R_{\dot{\alpha}_2}],\ldots,
\nabla^R_{\dot{\alpha}_n}]
\end{equation}
This method of defining the commutator {\it a priori} leads to an asymmetry
between $L$ and $R$. But we will later explain that in fact there is no such
asymmetry.

It remains to define $[x_L,\nabla^R_{\dot{\alpha}}]$. By definition $x_L\in {\cal L}_L$ is a linear combination 
of expressions of the form:
\begin{equation}\label{ProductOfNablaL}
   \nabla^L_{\alpha_1}\otimes\cdots\otimes\nabla^L_{\alpha_p}
\end{equation}
Moreover, it is a linear sum of nested commutators/anticommutators:
\begin{equation}
   x_L = [\nabla^L_{\alpha_1},\{\nabla^L_{\alpha_2},[\nabla^L_{\alpha_3},\ldots
\{\nabla^L_{\alpha_{p-1}},\nabla^L_{\alpha_p}\}\ldots]\}]
\end{equation}
(assuming that $x_L$ is odd). We define:
\begin{align}
\;&   \{\;\;\nabla^R_{\dot{\alpha}}\;,\;\;[\nabla^L_{\alpha_1},\{\nabla^L_{\alpha_2},
[\nabla^L_{\alpha_3},\ldots
\{\nabla^L_{\alpha_{p-1}},\nabla^L_{\alpha_p}\}\ldots]\}]\;\;\} \;=
\nonumber \\  
=\;&
[f_{\dot{\alpha}{\alpha_1}}{}^{[mn]}t^0_{[mn]},\{\nabla^L_{\alpha_2},
[\nabla^L_{\alpha_3},\ldots
\{\nabla^L_{\alpha_{p-1}},\nabla^L_{\alpha_p}\}\ldots ]\}]\; -
\nonumber \\  
\;& 
- \{\nabla^L_{\alpha_1}, [f_{\dot{\alpha}\alpha_2}{}^{[mn]}t^0_{[mn]},
[\nabla^L_{\alpha_3},\ldots
\{\nabla^L_{\alpha_{p-1}},\nabla^L_{\alpha_p}\}\ldots]\} \;+ 
\\   
\; & +
\ldots
\nonumber
\end{align}
We then do commute $t^0_{[mn]}$ with the remaining $\nabla^L$, and we are left with the sum 
of nested commutators of $p-1$ $\nabla^L$. We similarly define the commutator of $\nabla^R$
with any sequence of commutators of $\nabla^L$, not necessarily nested. There is 
a special case when $x_L=\nabla^L_{\alpha}$ ({\it i.e.} $p=1$); in this case we are left
with $f_{\dot{\alpha}\alpha}{}^{[mn]}t^0_{[mn]}\in {\bf g}_{\bar{0}}$. 

\subsubsection{Consistency}

\paragraph     {Jacobi identity}
We have to verify that the commutator which we defined satisfies the Jacobi
identity. There are two cases to verify:
\begin{enumerate}
\item $[[x_L,y_L], \nabla_{\dot{\alpha}}^R] = [x_L, [y_L,\nabla_{\dot{\alpha}}^R]]
\pm [[x_L,\nabla^R_{\dot{\alpha}}],y_L]$
\item $[t^0_{[mn]},[x_L,\nabla^R_{\dot{\alpha}}]] = [[t^0_{[mn]},x_L],\nabla^R_{\dot{\alpha}}] + [x_L,[t^0_{[mn]},\nabla^R_{\dot{\alpha}}]]$
\end{enumerate}
Both follow immediately from the definitions.

In particular, our definition does not depend on the choice of presentation of 
$x_L$ as a sum of nested commutators of $\nabla^L$. 

\paragraph     {Consistency with the quadratic relations}
There are two things to prove.

First, we have to prove:
\begin{equation}\label{ConsistencyRL}
\left[
 \langle{\cal R}_{Lm_1\ldots m_5}\rangle \;,\; \nabla^R_{\dot{\alpha}}
\right] \subset\;\; \langle{\cal R}_L\rangle
\end{equation}
This follows from:
\begin{equation}\label{RWithNabla}
   [{\cal R}_{Lm_1\ldots m_5} \;,\; \nabla^R_{\dot{\alpha}}] = 0
\end{equation}
The proof of (\ref{RWithNabla}) is formally indistinguishable from the verification of the 
Jacobi identity for $psl(4|4)$:
\begin{equation}
   [t^3_{\alpha},\{t^1_{\dot{\beta}},t^1_{\dot{\gamma}}\}] = 
[\{t^3_{\alpha},t^1_{\dot{\beta}}\},t^1_{\dot{\gamma}}] -
[t^1_{\dot{\beta}},\{t^3_{\alpha},t^1_{\dot{\gamma}}\}]
\end{equation}
Second, we have to prove:
\begin{equation}\label{xLWithRR}
   [x_L\;,\; {\cal R}_{Rm_1\ldots m_5} ] = 0
\end{equation}
Since $x_L$ is a nested commutator of $\nabla^L$, the Jacobi identity implies that it 
is enough to prove (\ref{xLWithRR}) for $x_L = \nabla^L_{\alpha}$. For any pure spinor $\lambda^{\dot{\alpha}}_R$ we should 
prove that $[\nabla^L_{\alpha},\{\lambda_R,\lambda_R\}]=0$, where $\lambda_R = \lambda_R^{\dot{\alpha}} \nabla^R_{\dot{\alpha}}$. Indeed:
\begin{equation}
   [\nabla^L_{\alpha}, \{\lambda_R,\lambda_R\}] = 2 [\{\nabla^L_{\alpha},\lambda_R\},\lambda_R] = 2 \lambda_R^{\dot{\gamma}}\lambda_R^{\dot{\beta}}f_{\alpha\dot{\beta}}{}^{[mn]}f_{[mn]\dot{\gamma}}{}^{\dot{\alpha}} \nabla_{\dot{\alpha}}^R
\end{equation}
This is zero since by the $psl(4|4)$ Jacobi identity 
$f_{\alpha\dot{\beta}}{}^{[mn]}f_{[mn]\dot{\gamma}}{}^{\dot{\alpha}}=-2f_{\dot{\beta}\dot{\gamma}}{}^mf_{m\alpha}{}^{\dot{\alpha}}$, and  $\lambda_R^{\dot{\beta}}\lambda_R^{\dot{\gamma}}f_{\dot{\beta}\dot{\gamma}}{}^m = 0$ because $\lambda_R$ is a pure 
spinor.

\paragraph     {Commutator of two primitive elements is a primitive element}
Consider a commutator $[x_L,x_R]$ where $x_L\in {\cal L}_L$ and $x_R\in {\cal L}_R$. We have defined
${\cal L}_L\subset A_L$ and ${\cal L}_R\subset A_R$ as the subsets of primitive elements. We have to 
prove that $[x_L,x_R]$ is either an element of ${\bf g}_{\bar{0}}$ or belongs to either $A_L$ or 
$A_R$ and {\em is primitive}. If $\mbox{deg }x_L \geq \mbox{deg }x_R$, then this follows 
immediately. If $\mbox{deg }x_L < \mbox{deg }x_R$, then the result of the commutator falls 
into $A_R$, and we have to show that it is a primitive element of $A_R$. Let us
proceed by induction in $\mbox{deg }x_R$. Start with $\mbox{deg }x_R = 1$, {\it i.e.}
$x_R = \nabla^R_{\dot{\alpha}}$. In this case either $\mbox{deg }x_L \geq \mbox{deg }x_R$ or $x_L\in {\bf g}_{\bar{0}}$. In both case 
the statement follows immediately. Now suppose that $\mbox{deg }x_R = n>1$. Then 
$x_R$ is a linear combination of elements of the form $[\nabla^R_{\dot{\alpha}}, y_R]$ 
where $\mbox{deg }y_R = n-1$. We have:
\begin{equation}
   [x_R\;,\;x_L] = [\nabla^R_{\dot{\alpha}},[y_R,x_L]] - [y_R,[\nabla^R_{\dot{\alpha}}, x_L]]
\end{equation}
Both terms on the right hand side are primitive by the assumption of the 
induction.

\subsubsection{L${\leftrightarrow}$R symmetry}
Our construction of the commutator $[{\cal L}_L,{\cal L}_R]$ is {\it a priori} asymmetric under
L${\leftrightarrow}$R. But in fact it is L${\leftrightarrow}$R symmetric. Notice that our construction implies
the basic commutation relations (\ref{NablaLNablaR}), (\ref{T0NablaL}), (\ref{T0NablaR}), (\ref{NablaLNablaL}), (\ref{NablaRNablaR}).
At the same time, it can be in fact {\em derived} from these relations.
These relations are L$\leftrightarrow$R symmetric, and our algebra ${\cal L}_{\rm tot}$ is characterized
by them, therefore ${\cal L}_{\rm tot}$ enjoys the L$\leftrightarrow$R symmetry.

\section{Generalization of the Lax pair}\label{sec:Lax}
The basic relations (\ref{PureSpinorConstraint}) imply:
\begin{align}
   \{\nabla_{\alpha}^L, \nabla_{\beta}^L\} = \;& f_{\alpha\beta}{}^m A_m^L
\\    
[\nabla_{\alpha}^L, A_m^L] = \; & f_{\alpha m}{}^{\dot{\beta}}P_{\dot{\beta}\gamma} W^{\gamma}_L
\label{NablaA}
\end{align}
and similar equations for the commutators of $\nabla^R_{\dot{\alpha}}$. \marginpar{$P_{\dot{\alpha}\alpha}$}
Here $P_{\dot{\beta}\gamma}$ is the constant bispinor corresponding to the background RR field 
strength.

We propose the following generalization of the Lax pair:
\begin{align}
   L_+  = \;& 
\left(
   {\partial\over \partial \tau^+} 
   + J_{0+}^{[mn]} t^0_{[mn]} 
\right) 
+ J_{3+}^{\alpha} \nabla^L_{\alpha} + J_{2+}^m A^L_m + (J_{1+})_{\alpha} W_L^{\alpha} +
 \nonumber \\   
& 
+ \lambda_L^{\alpha} w^L_{\beta +} \left(
   \{  \nabla^L_{\alpha} \;,\; W_L^{\beta}\}
   - f_{\alpha}{}^{\beta}{}^{[mn]} t^0_{[mn]}
\right)
\label{LPlus} \\     
L_- = \;&
\left(
   {\partial\over\partial\tau^-}
   + J_{0-}^{[mn]} t^0_{[mn]}
\right) 
+ J_{1-}^{\dot{\alpha}} \nabla^R_{\dot{\alpha}} 
+ J_{2-}^m A^R_m + (J_{3-})_{\dot{\alpha}} W_R^{\dot{\alpha}} +
\nonumber \\  
& 
+ \lambda_R^{\dot{\alpha}} w^R_{\dot{\beta}-}\left(
   \{ \nabla^R_{\dot{\alpha}} \;,\; W_R^{\dot{\beta}} \}
   - f_{\dot{\alpha}}{}^{\dot{\beta}}{}^{[mn]} t^0_{[mn]}
\right)
\label{LMinus}
\end{align}
Here the currents $J_{\pm}$ are defined in the same way as in the ``standard'' Lax 
pair (\ref{OrigLPlus}), (\ref{OrigLMinus}) except for $(J_{1+})_{\alpha}$ and $(J_{3-})_{\dot{\alpha}}$ which are related to the $J_{1+}^{\dot{\alpha}}$ 
and $J_{3-}^{\alpha}$ of (\ref{OrigLPlus}), (\ref{OrigLMinus}) by lowering the indices with $P_{\alpha\dot{\alpha}}$:
\begin{equation}
 (J_{1+})_{\alpha} = P_{\alpha\dot{\alpha}} J_{1+}^{\dot{\alpha}}\;,\quad
(J_{3-})_{\dot{\alpha}} = P_{\dot{\alpha}\alpha} J_{3-}^{\alpha}  
\end{equation}
The purpose of introducing this $P_{\alpha\dot{\alpha}}$ (which cancels with the $P_{\alpha\dot{\alpha}}$ of (\ref{NablaA})) is
to keep notations for generators as in the flat space YM algebra.

\vspace{10pt}

\noindent In the rest of this section we will verify various properties of 
the generalized Lax pair.

\subsection{Gauge invariance under $sp(2)\oplus sp(2)$}\label{sec:G0GaugeTransform}
The construction preserves the ${\bf g}_{\bar{0}}$ gauge invariance, in the following sense.
For a $\tau^{\pm}$-dependent``gauge parameter'' $\xi_0(\tau^+,\tau^-)$, consider the transformation
$\delta_{\xi_{\bar{0}}}$ acting on the currents (\ref{Jvsg}) and ghosts $w_{\pm}$, $\lambda_L,\lambda_R$ as follows:
\begin{align}
   \delta_{\xi_{\bar{0}}} g =\;& \xi_{\bar{0}} g
\\  
\delta_{\xi_{\bar{0}}} \lambda_{L|R} =\;& [\xi_{\bar{0}},\;\lambda_{L|R}]
\\  
\delta_{\xi_{\bar{0}}} w_{\pm} =\;& [\xi_{\bar{0}},\;w_{\pm}]
\end{align}
It results in the ``covariant'' transformation of $L_{\pm}$, {\it i.e.}: 
\begin{equation}
   \delta_{\xi_0}L_{\pm} = [\xi_0,\;L_{\pm}]\quad \mbox{ \small for }\xi_0\in {\bf g}_{\bar{0}}
\end{equation}

\subsection{Zero curvature equations}
The proof of $[L_+,L_-]=0$ is straightforward using the equations of motion.
The list of equations of motion can be found {\it e.g.} in Section 2.2 of \cite{Mikhailov:2007mr}. 
When $[L_+,L_-] = 0$ is verified in \cite{Vallilo:2003nx} for (\ref{LoopLPlus}) and (\ref{LoopLMinus}), only some (but not 
all) of the commutation relations of the twisted loop superalgebra are used. 
The definition of ${\cal L}_{\rm tot}$ is such that those commutation relations which are 
really needed to verify $[L_+,L_-]=0$ are identical to the commutations 
relations in the twisted loop algebra. The following observations are useful:

\subsubsection{Grading of ${\cal L}_{\rm tot}$}
\marginpar{$\mbox{deg}$}
The algebra ${\cal L}_{\rm tot}$ is ${\bf Z}$-graded, with $\mbox{deg}(\nabla^L_{\alpha}) = 1$ and $\mbox{deg}(\nabla^R_{\dot{\alpha}}) = -1$. 
Obviously, all elements of ${\cal L}_L\subset{\cal L}_{\rm tot}$ are of positive grade, while all 
elements of ${\cal L}_R\subset{\cal L}_{\rm tot}$ are of negative grade. Let ${\cal L}_{\rm tot}^{[m,n]}$ denote the subspace of
${\cal L}_{\rm tot}$ consisting of elements $x$ such that $m\leq \mbox{deg}\; x \leq n$. 
\marginpar{ ${\cal L}_{\rm tot}^{[m,n]}$  $L{\bf g}^{[m,n]}$}
The twisted loop 
algebra $L{\bf g}$ also has grading, $\mbox{deg }z = -1$, with similar notations $L{\bf g}^{[m,n]}$.

\subsubsection{Structure of ${\cal L}_{\rm tot}^{[-3,3]}$}
Notice that ${\cal L}_{\rm tot}^{[1,3]}$, as a linear space, 
coincides with ${\bf g}_{\bar{3}} + {\bf g}_{\bar{2}} + {\bf g}_{\bar{1}}$. Namely, $\nabla^L_{\alpha}$ corresponds to $t^3_{\alpha}$, then 
$\{\nabla_{\alpha}^L,\;\nabla_{\beta}^L\} = f_{\alpha\beta}{}^mA_m^L$ and $A^L_m$ corresponds to $t_m^2$, then 
$[\nabla^L_{\alpha},A_m^L]= f_{\alpha m}{}^{\dot{\alpha}}P_{\dot{\alpha}\beta}W^{\beta}$ and $P_{\dot{\alpha}\beta}W^{\beta}$ corresponds to $t^1_{\dot{\alpha}}$. Consider now ${\cal L}_{\rm tot}^{[-3,3]}$. 
Observation:
\begin{itemize}
\item As a linear space ${\cal L}_{\rm tot}^{[-3,3]}$ coincides with the twisted loop algebra $L{\bf g}$ in 
degrees from $-3$ to $3$
\item The commutators of elements of ${\cal L}_{\rm tot}^{[-3,3]}$  coincide with the commutators of 
the corresponding elements of the twisted loop algebra, but only as long as 
they do not lead outside the degree range $[-3,3]$
\end{itemize}

\subsubsection{Commutator of ${\cal L}_{\rm tot}^{[4,4]}$ with ${\cal L}_{\rm tot}^{[-3,-1]}$}\label{sec:StructureOfL44}
As a linear space ${\cal L}_{\rm tot}^{[4,4]}$ is a direct sum of ${\bf g}_{\bar{0}}$ and some complementary 
space ${\cal N}_L$:
\begin{equation}\label{DecompositionOfL44}
   {\cal L}_{\rm tot}^{[4,4]} = {\bf g}_{\bar{0}} \;\;\oplus \;\; {\cal N}_L
\end{equation}
As a linear space ${\cal N}_L \simeq ({\bf C}^5\otimes {\bf C}^5)$; it is generated by the elements of the
form $[A^L_m,A^L_n]$ for $m\in \{0,\ldots,4\}$ and $n\in \{5,\ldots,9\}$. 
\begin{itemize}
\item The commutator of the 
   elements of $\cal N$ with the elements of ${\cal L}_{\rm tot}^{[-3,-1]}$ is zero. 
\item The commutators  of ${\bf g}_{\bar{0}}\subset {\cal L}_{\rm tot}^{[4,4]}$ with the elements of ${\cal L}_{\rm tot}^{[-3,-1]}$ 
   can be described by identifying, as a linear space, ${\cal L}_{\rm tot}^{[-3,-1]}$ with $L{\bf g}^{[-3,-1]}$.
   The subspace ${\bf g}_{\bar{0}}\subset {\cal L}_{\rm tot}^{[4,4]}$ can be identified as a linear space with $L{\bf g}^{[4,4]}$. 
   With these identifications, the commutator of an element of ${\cal L}_{\rm tot}^{[-3,-1]}$ with 
   an element of  ${\bf g}_{\bar{0}}\subset {\cal L}_{\rm tot}^{[4,4]}$ is the same as the commutator of the 
   corresponding element of  $L{\bf g}^{[-3,-1]}$ with the corresponding element of 
   $L{\bf g}^{[4,4]}$. It falls into $L{\bf g}^{[1,3]}\simeq {\cal L}_{\rm tot}^{[1,3]}$.
\end{itemize}

\subsection{Gauge transformation of $w_{\pm}$}
The definition of $L_{\pm}$ is correct w.r.to the gauge transformations of $w_{\pm}$:
\begin{equation}
      \delta_{\Lambda} w_{\alpha +} = \Gamma_{\alpha\beta}^m\Lambda_{m +} \lambda^{\beta}
\end{equation}
This follows from the existence $F^L_{mn}$ such that $\{\nabla_{\alpha}^L\;,\; W^{\beta}_L\} = F^L_{mn}(\Gamma^{mn})_{\alpha}^{\beta}$ ---
see Appendix B of \cite{Mafra:2009wq}.

\subsection{BRST transformation}
 We still have the standard BRST transformation rule:
\begin{equation}\label{BRSTCovariance}
   Q_{BRST}L_{\pm} \;=\; 
\left[
   L_{\pm}\;,\;\left(
      \lambda_L^{\alpha}\nabla^L_{\alpha} + 
      \lambda_R^{\dot{\alpha}} \nabla^R_{\dot{\alpha}}
   \right)
\right]
\end{equation}

\subsection{Comment about the coupling of ghosts}
Notice that the coupling to $\lambda_L w_+$ and $\lambda_R w_-$ involves the full $so(10)$ Lorentz
currents, not only the $so(1,4)\oplus so(5)$ part. Indeed, the term 
$\lambda_L^{\alpha} w^L_{\beta +} 
   \{  \nabla^L_{\alpha} \;,\; W_L^{\beta}\}$ in (\ref{LPlus}) can be split into two parts according to (\ref{DecompositionOfL44}):
\begin{equation}
   \lambda_L^{\alpha} w^L_{\beta +} 
   \{  \nabla^L_{\alpha} \;,\; W_L^{\beta}\} = 
\lambda_L^{\alpha} w^L_{\beta +} 
   \left(
      \left.
         \{  \nabla^L_{\alpha} \;,\; W_L^{\beta}\}
      \right|_{{\bf g}_{\bar{0}}} + 
      \left.
         \{  \nabla^L_{\alpha} \;,\; W_L^{\beta}\}
      \right|_{{\cal N}_L}
\right)
\end{equation}
Since the commutator of ${\cal L}_R$ with ${\cal N}_L$ is zero, the zero curvature equation
$[L_+,L_-]=0$ would hold even if we drop the coupling to $\left.
         \{  \nabla^L_{\alpha} \;,\; W_L^{\beta}\}
      \right|_{{\cal N}_L}$. The
requirement of ${\bf g}_{\bar{0}}$-gauge invariance, as well as the invariance under the gauge 
transformation of $w_{\pm}$, would be also satisfied. The only thing that would 
break is the BRST covariance (\ref{BRSTCovariance}). Indeed, for $L_{\pm}$ to be BRST-covariant, 
the terms of $L_+$ (and similarly $L_-$) should satisfy the chain of identities:
\begin{align}
\left[ \epsilon \lambda_L^{\alpha}\nabla^L_{\alpha}\;,\;  \left(
   {\partial\over \partial \tau^+} 
   + \left(
      J_{0+}^{[mn]} - \lambda_L^{\alpha} w_{\beta +}^L f_{\alpha}{}^{\beta [mn]}
   \right) t^0_{[mn]}
\right) 
\right] = \;& 
\epsilon Q_LJ_{3+}^{\alpha}\nabla^L_{\alpha}
\\    
\left[ 
   \epsilon \lambda_L^{\alpha}\nabla^L_{\alpha}\;,\;  
   J^{\beta}_{3+}\nabla^L_{\beta}
\right] = \;&   
\epsilon Q_LJ_{2+}^{m}A_m^L
\\   
\left[
   \epsilon \lambda_L^{\alpha}\nabla^L_{\alpha}\;,\;  
   J_{2+}^m A^L_m
\right] = \;&   
\epsilon Q_L(J_{1+})_{\alpha} W_L^{\alpha}
\\    
\left[
   \epsilon \lambda_L^{\alpha}\nabla^L_{\alpha}\;,\;  
   (J_{1+})_{\beta}W^{\beta}_L
\right] = \;&
\lambda_L^{\alpha}\epsilon Q_L w^L_{\beta +}\times 
\nonumber\\    
& \times \{\nabla^L_{\alpha}\;,\;W_L^{\beta}\}
\label{LinkWithQw}
\\    
\left[
   \epsilon \lambda_L^{\gamma}\nabla^L_{\gamma}\;,\;  
   \lambda_L^{\alpha} w^L_{\beta +} 
   \{  \nabla^L_{\alpha} \;,\; W_L^{\beta}\}
\right] = \;& 0
\end{align}
If we drop the coupling to $\left.
         \{  \nabla^L_{\alpha} \;,\; W_L^{\beta}\}
      \right|_{{\cal N}_L}$, or modify its coefficient, then
Eq. (\ref{LinkWithQw}) will break. 

Generally speaking, BRST invariance should imply the zero curvature conditions,
along the lines of \cite{Puletti:2008ym}. But the presented example shows that the zero curvature
condition does not necessarily imply the BRST invariance. 

Notice that ${\cal N}_L$ generates an ideal $I({\cal N}_L)$ of ${\cal L}_{\rm tot}$ which is a subset of ${\cal L}_L$:
\begin{equation}
   I({\cal N}_L)\subset {\cal L}_L
\end{equation}
and similarly ${\cal N}_R$. It seems plausible that the factoralgebra 
${\cal L}_{\rm tot}/(I({\cal N}_L)+I({\cal N}_R))$ is $L{\bf g}$, but we have not looked into that.

\section{Open questions}\label{sec:OpenQuestions}
Given the Lax pair $L_+,L_-$, of the form:
\begin{equation}
L_{\pm} = {\partial\over\partial\tau^{\pm}} + A_{\pm}   
\end{equation}
one can construct the transfer-matrix:
\begin{equation}
   T(\tau_{\rm fin},\tau_{\rm in}) = P\;\mbox{exp}\left( 
      - \int_{\tau_{\rm in}}^{\tau_{\rm fin}} (A_+d\tau^+ + A_-d\tau^-)
   \right)
\end{equation}
This is not ${\bf g}_{\bar{0}}$-gauge invariant. The variation of $T(\tau_{\rm fin},\tau_{\rm in})$ under the gauge 
transformation of Section \ref{sec:G0GaugeTransform} is:
\begin{equation}
   \delta_{\xi} T(\tau_{\rm fin},\tau_{\rm in}) = 
\xi(\tau_{\rm fin})T(\tau_{\rm fin},\tau_{\rm in})
- T(\tau_{\rm fin},\tau_{\rm in})\xi(\tau_{\rm in})
\end{equation}
One could say that the transfer matrix is gauge invariant ``up to the boundary
terms''. These boundary terms could be cancelled by considering the trace of 
the transfer matrix over the closed contour. Alternatively, one could imagine 
some networks of Wilson lines corresponding to different representations, with 
${\bf g}_{\bar{0}}$-invariant vertices. In any case, there is a potential difficulty. It is not
clear for which representations of ${\cal L}_{\rm tot}$ the trace of $T$ is well-defined. It may
be the case, that the trace is only well-defined in those representations which
are representations of the twisted loop algebra. If this is true, then our 
construction does not give any new conserved charges. 

In any case, it would be interesting to understand the representations 
of ${\cal L}_{\rm tot}$. Moreover, even the representations of the Yang-Mills algebra ${\cal L}_L$ are 
not well-understood. Physically, what do they correspond to?

\section*{Acknowledgments}
This work was supported in part by the Ministry of Education and Science of 
the Russian Federation under the project 14.740.11.0347 ``Integrable and 
algebro-geometric structures in string theory and quantum field theory'', and 
in part by the RFFI grant 10-02-01315 
``String theory and integrable systems''.


\def\cprime{$'$} \def\cprime{$'$}
\providecommand{\href}[2]{#2}\begingroup\raggedright\endgroup

\end{document}